# From surface roughness to crater formation in a 2D multi-scale simulation of ultrashort pulse laser ablation

N. Thomae[1,†], M. Stabroth[2,3,†], J. Vollmann[1], M. Döring[2,3,*], D. Redka[1,4], H.P. Huber[1,4], M. Schmidt[2,3]

[1]Munich University of Applied Sciences HM, Munich, Germany
[2]Friedrich-Alexander-Universität Erlangen-Nürnberg, Erlangen, Germany
[3]Friedrich-Alexander-Universität Erlangen-Nürnberg, Erlangen Graduate School in Advanced Optical Technologies (SAOT), Erlangen, Germany
[4]New Technologies Research Center, University of West Bohemia, Plzen CZ-30100, Czech Republic
[†]These authors contributed equally to this work
*email of corresponding author: markus.doering@fau.de

## Abstract

Surface roughness plays a critical role in ultrashort pulse laser ablation, particularly for industrial applications using burst mode operations, multi-pulse laser processing, and the generation of laser-induced periodic surface structures. Hence, we address the impact of surface roughness on the resulting laser ablation topography, comparing predictions from a simulation model to experimental results. We present a comprehensive multi-scale simulation framework that first employs finite-difference-time-domain simulations for calculating the surface fluence distribution on a rough surface measured by atomic-force-microscopy followed by the two-temperature model coupled with hydrodynamic/solid mechanics simulation for the initial material heating. Lastly, a computational fluid dynamics model for material relaxation and fluid flow is developed and employed. Final state results of aluminum and AISI 304 stainless steel simulations demonstrated alignment with established ablation models and crater dimension prediction. Notably, Al exhibited significant optical scattering effects due to initial surface roughness of 15 nm - being 70 times below the laser wavelength -leading to localized, selective ablation processes and substantially altered crater topography compared to idealized conditions. Contrary, AISI 304 with $R_q$ surface roughness of 2 nm showed no difference. Hence, we highlight the necessity of incorporating realistic, material-specific surface roughness values into large-scale ablation simulations. Furthermore, the induced local fluence variations demonstrated the inadequacy of neglecting lateral heat transport effects in this context.

## Introduction

Ultrashort pulse (USP) laser ablation is a pivotal technique used in industrial and scientific applications as it enables the modification of surfaces across macroscopic to micro-/nanoscopic scales, functionalizing mechanical, electrical, chemical, and optical properties [1–3]. For predicting and optimizing ablation surface topography, simulating the ablation process serves as a critical tool to enhance the efficiency and accuracy of laser-based manufacturing and processing. This, however, poses an ongoing challenge due to its complex multi-physics and multi-scale nature. Hence, diverse computational methods such as molecular dynamics [4], hydrodynamics (HD) [5,6] and solid mechanics (SM) [7], all including the two-temperature model (TTM), have been utilized to address distinct aspects of the ablation process.

USP laser pulses in the fs and ps regime lead to energy being confined to the nm scale below the surface. Within few ps, depending on the material-specific strength of the electron-phonon coupling [8], thermal equilibrium between the electrons and the lattice is reached, creating a melted layer of nm thickness and high pressure build-up below the surface in the GPa regime [9,10]. After roughly 15 ps the high pressure induces fluence-dependent mechanisms like spallation and phase explosion, which lead to material ablation [11]. Lastly, on the ns-timescale and lasting up to µs [12], material relaxation and fluid reorganization due to driving forces like (thermo)capillarity [13] or Marangoni convection [14] eventually determine the crater topography.

For idealized flat surfaces, the problem is well-handled. Despite the efficacy of current simulations in trend prediction, they fall short of accurately forecasting experimental topographies due to the oversimplification of surface conditions, neglecting surface roughness, impurities, and material defects. Especially in context of industrial applications with burst pulses and multi-pulse ablation, these surface alterations subsequently influence the ablation process by scattering incident electrical fields creating surface waves [15], inducing local field enhancements and subsequently increasing the local optical absorption [16]. Furthermore, these scattering effects play an essential role in forming self-organizing laser-induced periodic surface structures (LIPSS) [17,18].

Therefore, we demonstrate a multi-scale approach coupling physical domains from an optical simulation using finite-difference-time-domain (FDTD) simulation of real surfaces with nm resolution to 1D TTM coupled with continuum mechanic (HD/ SM) (from here on referred to as only TTM) for simulating the material response which in turn, feeds into a compressible multi-phase computational fluid dynamics model with an included TTM (from here on referred to as only CFD) for describing material relaxation, phase transitions and fluid flow in 2D and on a longer time scale, thereby achieving a multi-scale and multi-physics framework to encapsulate the whole ablation process and predict final states fully. The final state observables, namely crater geometry, topography, and threshold fluences from the simulations, were compared with published experimental findings by Winter *et al.* [19] and established ablation models, with particular attention given to the differences observed between ideal and rough surfaces. Two different materials stainless steel (AISI 304) and aluminum (99.999 % purity) are compared due to their industrial importance but also significant difference in surface roughness despite the same polishing procedure with a $R_\text{q}$ value of 15 nm and 2 nm for Al and AISI 304, respectively.

## Materials and method
*Multi-scale simulation approach*

Our approach to resolving the problem of USP laser-matter interactions on real surfaces is sketched in Fig. 1. Initially, the (volume) heat source is determined by calculating the locally absorbed laser intensity by solving the Maxwell equations via the FDTD method for a real two-dimensional surface, obtained by Atomic Force Microscopy (AFM). These computations precisely determine the spatially modulated field enhancements primarily attributable to scattering [20], excitation of electromagnetic surface waves like surface plasmon polaritons (SPP) [21] and the resulting interference with the incident laser pulse. From this spatial fluence distribution, a mosaic of thermophysical properties and initial conditions is constructed for the CFD simulation domain by calculating a fluence catalogue via TTM simulation. This catalogue is compiled from solutions to TTM for discrete incident fluences. These solutions consist of the thermodynamic state variables (electron and lattice temperatures, pressure, phase, velocity, density) over a depth of 10 µm. The endpoint and thus the transition time of the TTM to CFD simulations is set at 1.575 ps, three times the pulse duration, ensuring that >99.99% of the pulse energy is absorbed, thereby obviating the need for a time-dependent heat source term for the CFD calculations. The CFD solver resolved the relaxation of the excited system from thermodynamic non-equilibrium within a two-dimensional domain. The chosen strategy enables the computation of optical phenomena induced by surface roughness, the solution of the TTM model under varying transport and optical properties, as well as two-dimensional hydrodynamic processes, including vaporization and hydrodynamic movements such as Marangoni flow via an integrated framework. The incorporated TTM approach assumes that lateral heat transfer is negligible due to the low lateral thermal gradient relative to the gradient normal to the surface. The maximal temperature gradient can be estimated laterally with the beam radius $w_0$ and vertically with optical penetration depth $d_\text{opt}$. This results in an estimation of the heat transport $\dot{Q}_\text{lateral} \propto 1/w_0 \propto 1/(10\ \text{µm})$, and $\dot{Q}_\text{vertical} \propto 1/d_\text{opt} \propto 1/(10\ \text{nm})$, with the beam radii $w_0$ and the optical penetration depth $d_\text{opt}$, resulting in a vertical heat transfer rate that is roughly a thousandfold bigger than the lateral heat transfer.

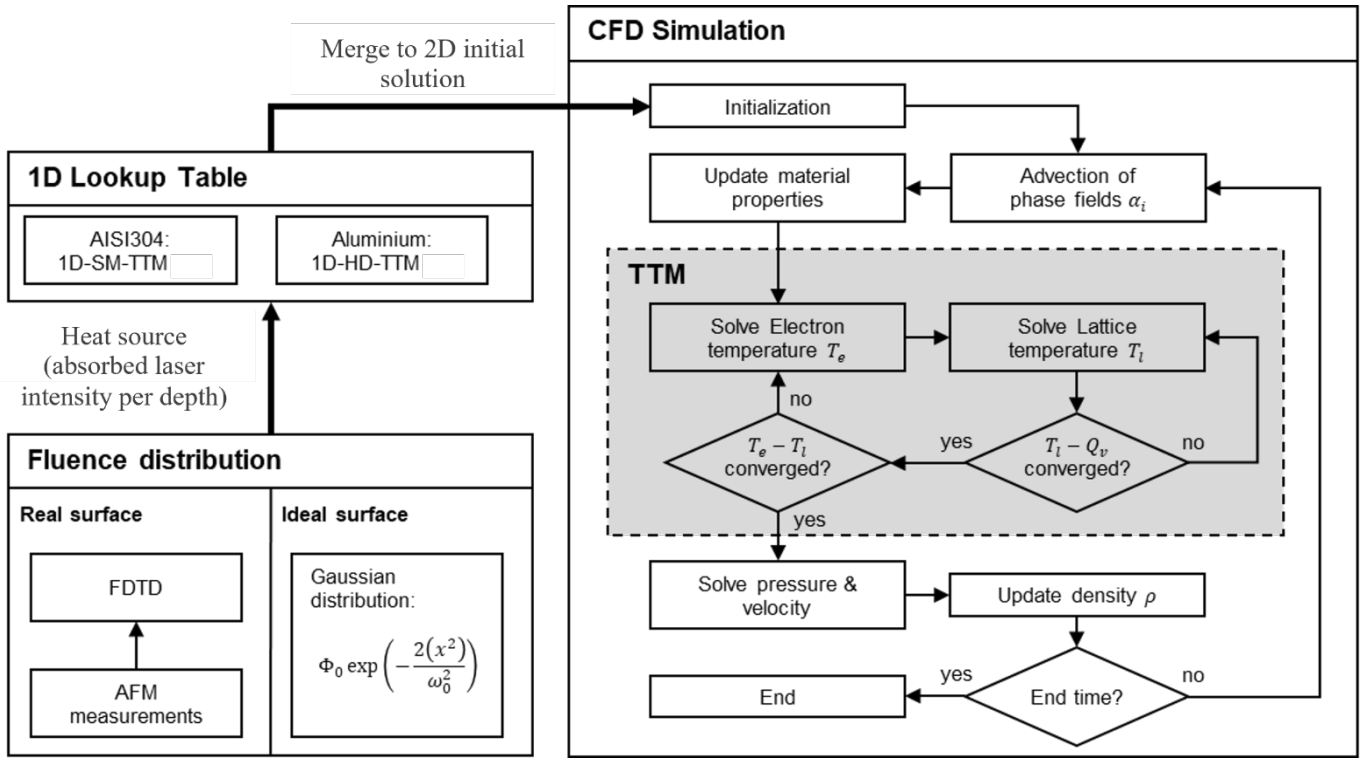

**Fig. 1** Schematic representation of multi-scale simulation approach. Starting with the fluence distribution from FDTD simulations of a rough surface or idealized Gaussian fluence distribution for an ideal surface into a 1D-lookup table for thermodynamical properties as pre-calculated by TTM simulation. A subsequent CFD model, including TTM, utilizes these derived thermodynamic properties, post 1.575 ps laser irradiation, to calculate material relaxation and predict final states.

*FDTD and surface measurement*
FDTD is a broadly applicable numerical method for computational electrodynamics [22]. The commercially available software Lumerical (Lumerical FDTD solver from Ansys optics, version 2021, R1.3) was utilized for the simulation. Initially, a Gaussian pulse in space and time, with a wavelength $\lambda$ of 1056 nm, pulse radius $w_0$ of 15 µm, and pulse duration $\tau_p$ of 525 fs with parallel polarization is implemented, adapted from previously published experimental results by Winter *et al.* [19].
The surface topography was implemented by extracting height profiles from AFM measurements and discretized in the software into the computational cells. The lateral resolution of the simulation grid was set to 25 nm, approximately 40 times smaller than the laser wavelength, to capture the relevant surface features influencing the intensity distribution. The vertical resolution was set to 1 nm to resolve the intensity distribution in depth via linear absorption, with a stability factor of 0.99 to fulfil the Courant condition [22]. The temporal resolution was set to dt = 0.003 fs. The boundaries of the simulation domain were implemented with perfectly matched layers with 8 layers. The lateral width of the simulation was set to $\pm 3w_0$, ensuring intensities of the Gaussian pulse remain below 0.0001% at the boundaries to minimize residual reflectance errors. Lastly, the duration of the simulation was set to $\pm 3\tau_p$ with the laser pulse intensity maximum at $t = 0$ s, ensuring minimal intensity loss of the Gaussian source of less than 0.01%.

The complex refractive indices were taken from literature for Al [23] and AISI304 [24]. Transient change of the dielectric function due to electron heating, as observed in pump-probe ellipsometry experiments for both metals [25], was not directly considered in the FDTD simulations. The simulation results were focused solely on determining the geometrical influence on the field distribution. For Al, the TTM simulation accounted for these transient changes, whereas for AISI 304, these changes were not included. This is discussed in more detail in the subsequent section.

To compute the field intensities, a spatial region was defined around the interface between the vacuum and the surface, where the electrical field components were calculated in the frequency domain and evaluated at the central wavelength. By extracting a matrix representing the refractive index within the same spatial region, a mask was defined, which was used to determine the field intensities at the grid points of the surface. An example of the simulation setup as well as comparison of the 2D approximation to a 3D simulation is given in the appendix.

Using the relative intensities ($I_{\text{real}}/I_{\text{ideal}}$), the local fluence of the real surface simulation can be scaled according to $\Phi_{\text{real}}(x) = \Phi_{\text{ideal}}(x) \cdot (I_{\text{real}}/I_{\text{ideal}})$, with the ideal Gaussian intensity distribution $\Phi_{\text{ideal}}(x) = \Phi_0 \exp(-2(x^2)/\omega_0^2)$. With this, the intensity distribution of the FDTD simulation can be normalized for any arbitrary chosen fluence $\phi_0$ based on the linearity of the intensity calculation.

For the study two industrially relevant metals were chosen, namely Al and AISI 304. To generate surface data of real, polished Al and AISI 304 samples, an AFM (JPK Instruments, NanoWizard 1, Germany) with Cantilever (OMCL AC-160TS Olympus, Japan) with tip radius 8 nm in tapping mode resulting in a lateral resolution of approx. 25 nm was utilized. Furthermore, the software Gwyddion was used with standard procedures for AFM measurements to remove measurement artefacts. The surfaces of the polished Al and the AISI 304 yielded a surface roughness averaged over an area of approximately 25 µm² of $S_q = 15$ nm ($S_a = 12$ nm) and $S_q = 2$ nm ($S_a = 1.4$ nm), respectively, with the same polishing procedure.

*1D-TTM-Continuum mechanic simulation*
According to the described multi-scale simulation approach, 1D-TTM simulations were conducted for Al and AISI 304, which serve as input for later CFD calculations.

The thermodynamical properties density, pressure, phase, electron-/ and ion temperature were calculated for distinct fluences $\phi_{0,\text{inc}}$ in steps of 0.5 mJ/cm² (incident fluences for Al: 0.5 mJ/cm² − 3000 mJ/cm² and absorbed fluences for AISI 304: $0.5 \cdot (0.5$ mJ/cm² − 280 mJ/cm²)) up to the time of 1.575 ps. This duration corresponds to three times the pulse length, ensuring that 99% of the laser pulse was absorbed. This time span represented the shortest possible interval for the 1D approximation. Boundary conditions were set to thermal insulators at all boundaries with 1 µm depth and an initial temperature of 300 K.

For Al, a TTM-HD simulation was employed, connecting the TTM calculation to a numerical hydrodynamic solver with a defined Equation of States (EOS) including temperature dependent thermodynamical and optical properties. Incident fluences were used as this model incorporates transient changes of the complex refractive index including changes in the dielectric function and density. The TTM-HD was experimentally validated prior to this publication and detailed information about the computation and material parameters can be found in [23]. Although TTM-HD includes the mechanisms of spallation and phase explosion, the 1.575 ps timeframe is insufficient to observe their influence [26].

The material response of AISI 304 is simulated with a TTM-SM model using the FEM software COMSOL (version 5.6) containing relevant EOS as HD simulations are not available. Computational details about this method can be found in a prior work [27] for calculations and verifications of Al simulations. Here, this methodology was implemented for AISI 304 with numerically interpolated thermodynamical and optical parameters taken from [24] and the EOS from the SESAM database [28], which is shown in the appendix. Contrary to the TTM-HD model for Al, transient changes in the complex refractive index were not considered. Therefore, absorbed fluences were directly applied. Based on experimental pump-probe ellipsometry results [25], the changes for the real part are negligible and the imaginary part decreases below 15% during pulse irradiation. Despite these variations, this approach still provides a qualitatively accurate representation.

*2D-TTM-CFD Simulation*

A compressible multi-phase solver, including the TTM, is implemented in OpenFOAM utilizing the Volume of fluid method (VOF), thereby completing the envisioned multi-scale simulation toolchain. This CFD model focuses on material relaxation, fluid flow, phase transitions and the resulting ablation crater topologies. The underlying phase transitions during USP laser processing necessitate the ability to simulate an arbitrary number of phases $\alpha_i = [0,1]$ and incorporate the TTM for describing the temporal behavior of the heat source. OpenFOAM utilizes the finite volume method, which has a high accuracy for discontinuities [29] and conservation of the flows [30]. The conservation equations for mass and momentum [31]

$$\frac{\partial \rho}{\partial t} + \nabla \cdot (\rho \mathbf{U}) = 0 \tag{1}$$

$$\frac{\partial (\rho \mathbf{U})}{\partial t} + \nabla \cdot (\rho \mathbf{U}\mathbf{U}) = -\nabla p + \nabla \mathbf{\tau} + \rho \mathbf{g} + \mathbf{F_S} + \mathbf{F_D} \tag{2}$$

are implemented. With the stress tensor of a Newtonian fluid

$$\mathbf{\tau} = \mu[\nabla \mathbf{U} + (\nabla \mathbf{U})^T] - \frac{2}{3}\mu(\nabla \cdot \mathbf{U}). \tag{3}$$

Whereby, $\rho$ is density, $t$ time, $\mathbf{U}$ the velocity, $p$ pressure, $g$ the vector of gravitational acceleration and $\mu$ the dynamic viscosity. The vectors $\mathbf{F}$ represent the source/sink terms in the equation for conservation of momentum and are included to model solidification and the forces acting on the free surfaces. $\mathbf{F_D}$ is used to model the solidification interval. It acts as a damping force and describes the flow through a porous medium according to Darcy's law and the Carman-Kozeny equation to replicate the behavior of solidifying material by restricting the movement of the material [32,33]. $\mathbf{F_S}$ models the remaining forces acting on the surfaces

$$\mathbf{F_S} = \left\{ \sigma \kappa \mathbf{N} + \frac{d\sigma}{dT}[\nabla T - \mathbf{N}(\mathbf{N} \cdot \nabla T)] + \mathbf{N} \cdot P_v \right\} |\nabla \alpha_{\text{sum}}| \frac{2\rho}{\rho_l + \rho_v}. \tag{4}$$

It comprises the recoil pressure $P_v$ due to evaporation, as well as the normal and tangential components of the surface tension. The tangential component of the surface tension $\sigma$ thereby represents the thermocapillary force driving the Marangoni convection [34,35]. The geometric information regarding the surface is provided by the normal vector $\mathbf{N} = \nabla \alpha / |\nabla \alpha|$ and the curvature $\kappa = -(\nabla \cdot \mathbf{N})$ of the surface. The gradient of the phase fractions of the sum of solid and liquid materials $\alpha_{\text{sum}}$ is used to direct the forces on the free surface separating liquid material and metal vapor/environmental air. The recoil pressure is given by the Clausius-Clapeyron equation; it considers the condensation rate depending on the ambient pressure by choosing an appropriate re-condensation factor $\beta$ of 0.55 [36,37]

$$P_v = \beta P_0 \exp\left[\frac{L_v M (T - T_v)}{R T T_v}\right]. \tag{5}$$

It includes the ambient pressure $P_0$, the enthalpy of vaporization $L_v$, the molar mass $M$, the boiling temperature $T_v$ and the universal gas constant $R$.

As thermal equilibrium between electrons and phonons is not established at the transition point from TTM to CFD at 1.575 ps for both materials, the TTM is also incorporated into the CFD, given by [38]

$$\frac{\partial}{\partial t}(\rho C_e T_e) = \nabla \cdot (k_e \nabla T_e) - G(T_e - T_l), \tag{6}$$

$$\frac{D}{Dt}(\rho C_l T_l) = G(T_e - T_l) + Q_{\text{PC}}. \tag{7}$$

With the temperatures of the electrons $T_e$ and the lattice $T_l$, the (volumetric) specific heat capacities, $C_e$, $C_l$, the heat capacity of the electrons $k_e$, as well as the electron-phonon coupling constant $G$, which characterizes the energy exchange between the two systems. $Q_{\text{PC}}$ describes the heat flows due to phase transitions and is described by

$$Q_{\text{PC}} = -L_f \left[ \frac{\partial}{\partial t}(\rho \alpha) + \nabla \cdot (\rho \alpha \mathbf{U}) \right] - L_v \dot{m}_v \tag{8}$$

for melting and solidification [32]. $L_f$ therein is the enthalpy of fusion. The TTM is calculated using two loops whose termination criteria are linked to the temperature change from the previous iteration. The inner loop solves the lattice temperature and the phase transition enthalpies, whereas the outer loop solves the equations of the electron and lattice temperatures. The VOF method is used for advection and tracking of the free surfaces. It solves the transport equations for each phase. The local fluid properties are averaged for a single mixture over the respective phase fraction accordingly [39]. The transport equation of the phase fractions with the compressibility $\psi$

$$\frac{\partial \alpha}{\partial t} + \nabla \cdot (\mathbf{U}\alpha) - \dot{\alpha}_{\text{pc}} = \alpha \nabla \cdot \mathbf{U} + \alpha(1-\alpha)\left(\frac{\psi^v}{\rho^v} - \frac{\psi^l}{\rho^l}\right)\frac{Dp}{Dt} \tag{9}$$

is described in [40,41]. The phase transition rate $\dot{\alpha}_{\text{pc}}$ due to bi-directional phase transition can be calculated [39], for example, for the vaporization of the liquid phase as

$$\dot{\alpha}_{\text{pc}} = -\dot{m}_\alpha \cdot \left(\frac{1}{\rho_l} - \alpha_l \cdot \left(\frac{1}{\rho_l} - \frac{1}{\rho_v}\right)\right). \tag{10}$$

The implemented model includes five separate phases: solid, resolidified, liquid, vapor and ambient air. For the phase transitions, mainly thermodynamic considerations are included in the model, kinetic aspects like nucleation meditated phase transitions behavior resulting in superheating and undercooling, beside the Clausius-Clapeyron equation, are not considered. The domain size was chosen depending on the simulated surface and material. The domain had a width of 30 μm for the real surface case and 15 μm for symmetric simulations of the ideal surface. The height of the domain was chosen so the temperature field and the ejected vapor could spread sufficiently. Sizes between 700 nm and 4 μm were chosen for this purpose. The static mesh consists of up to approx. 250,000 hexahedra with a lateral extent of 25 nm and a vertical extent of up to 1 nm around the free surface of the material. The initial time step size is $10^{-24}$ s and increases to a maximum of $10^{-15}$ s given by the compressible CFL number of $10^{-2}$ [42]. Running on 32 cores of an AMD EPYC High Performance Workstation (AMD Threadripper 3970x; 128 GB DDR4-2666 ECC), a simulation runs for approx. 4 days until the material is completely resolidified.

The material models for aluminum and AISI 304 were predicated on the utilization of the EOS spanning a broad temperature and pressure range, employing the Lomonosov EOS [43] and Sesame EOS [28], respectively. These models describe the free energy of the system as a function of density and temperature. They enable the derivation of various thermodynamic quantities - such as pressure, heat capacities, bulk modulus, and phase transitions – dependent on density and temperature parameters. The fundamental variables of the implemented model are pressure ($p$) and velocity ($\mathbf{U}$), and the temperatures of the electrons $T_e$ and lattice $T_l$. This necessitates the preliminary computation of density from pressure $p(T,\rho) \rightarrow \rho(T,p)$. For this, a 1024x1024 density matrix $\rho(T,p)$ was constructed from the aforementioned EOS. From the initial pressure and temperature, the density in each cell is interpolated from this density matrix. After the pressure field is updated, the density is also recalibrated for each step is based on this density matrix. This recalibration is essential as it enables updating other material parameters based on temperature and density, for instance, the specific heat capacity $C(T,\rho)$. The surface tension $\sigma$ is interpolated between melting temperature and the critical temperature. A comprehensive overview of the relevant material properties is compiled in Table 1.

**Tab. 1** Temperature-dependent thermophysical properties of Al and AISI 304. The phase boundaries of the EoS for Al are shown in [27] and for AISI 304 depicted in the appendix.

| Thermophysical Property | Aluminum | AISI 304 |
| --- | --- | --- |
| Density, $\rho$ | EOS based on [43] | EOS based on [28] |
| Specific heat, $C$ | EOS based on [43] | [24] |
| Thermal conductivity of electrons, $\kappa_e$ | [23] | [24] |
| Melting temperature, $T_m$ | [44] | EOS based on [28] |
| Vaporization temperature, $T_v$ | [45] | [46] |
| Electron phonon coupling factor, $G$ | [23] | [24] |
| Dynamic viscosity, $\mu$ | see Appendix (JMatPro) | see Appendix (JMatPro) |
| Surface tension, $\sigma$ | [47] | 2.6552 N/m (JMatpro) |
| Compressibility, $\psi$ | EOS based on [43] | EOS based on [28] |

## Results

The laser parameters (in simulation and experiment) under investigation were set to a pulse duration of $\tau_p = 525$ fs (FWHM), a wavelength $\lambda$ of 1056 nm and a beam radius $w_0$ of 15 µm ($1/e^2$-definition). Simulations across the toolchain were conducted for Al and AISI 304 and their results are compared to experimental data.

*Time-Resolved Ablation Dynamics by TTM-CFD*

Fig. 2 depicts the ablation dynamic for an ideally flat surface for the fluences of 0.885 Jcm$^{-2}$ for Al and of 0.195 Jcm$^{-2}$ for AISI 304 (1.5 times the ablation threshold). The thermodynamic phases are marked red and yellow for vapor and liquid, as well as dark and bright blue for solid and resolidified material. Delay times $t$ are denoted relative to the temporal peak of the Gaussian pulse.

At 2.075 ps, Al is vaporized to a depth of approximately 40 nm in the crater center ($x = 0$), with a radius of 6 µm, while melting occurs up to a depth of 100 nm and a radius of 12 µm. Both phases show a parabolic profile. At 11.575 ps, the vapor phase deepened to 50 nm, while the previously melted region is already partially resolidified, especially in the center and at the crater's edge, with a resolidification velocity of 5.6 km/s.

At 101.575 ps and 201.575 ps, more material is accumulated at the crater edges, indicating lateral melt flow towards the edges. During this period, resolidification is nearly complete, forming a crater with a radius of 7 µm and a depth of 70 nm. Concurrently, the vapor phase expands inhomogeneously above the surface at an approximate velocity of 1.1 km/s. The vapor retains a high density, only four times the volume of the original solid material pre-irradiation.

AISI 304, at 2.075 ps, is vaporized to a depth of 25 nm and a radius of 7 µm and molten to a depth of 40 nm and a radius of 13 µm. By 6.575 ps, the liquid phase is almost entirely resolidified with a resolidification velocity of 6.25 km/s, showing no more substantial change in depth or diameter. At 51.575 ps, the vapor phase is ejected with a velocity of 3.3 km/s. Compared to Al, the vapor plume is more uniform, showing no increase in ejection velocity near the crater's edge.

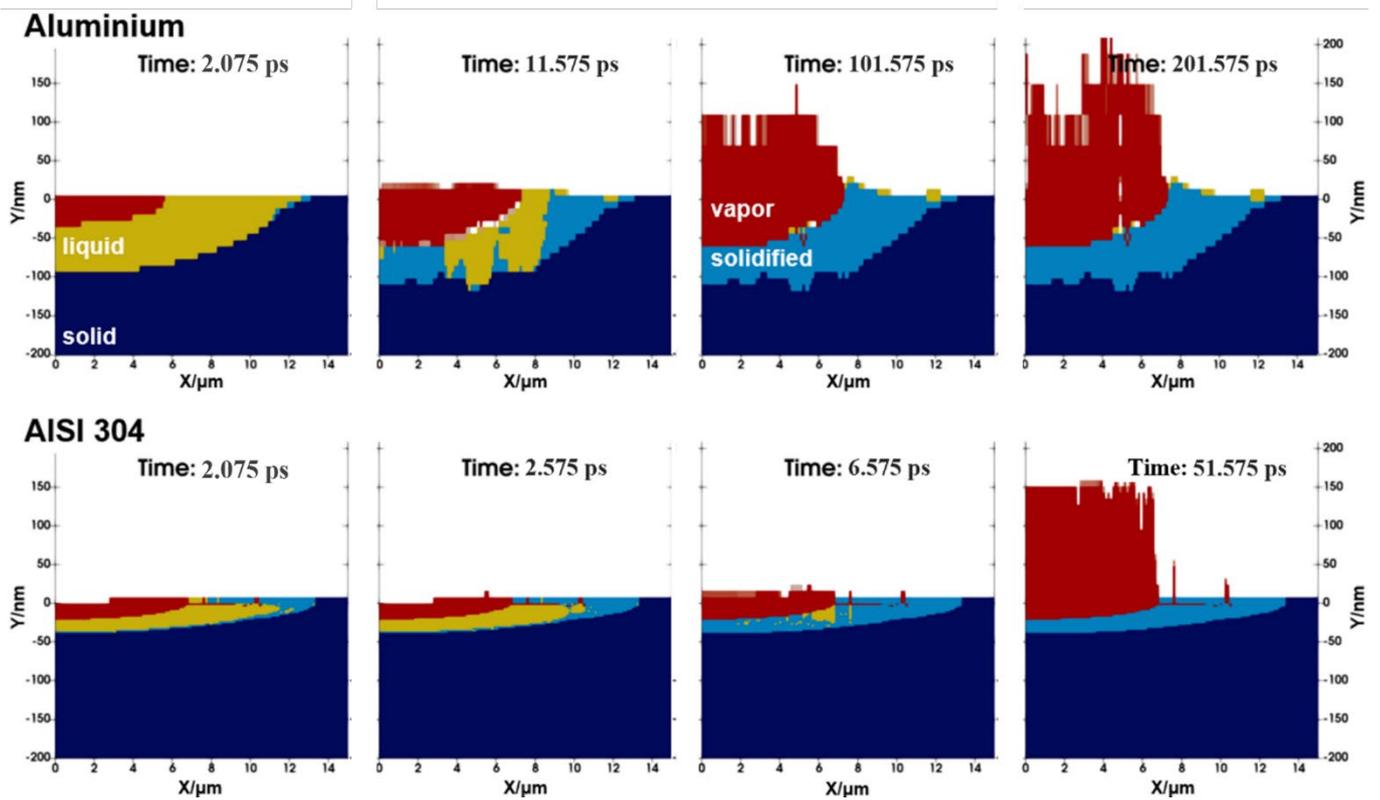

**Fig. 2** Snapshots from a CFD simulation depicting the phases in Al and AISI 304 at varying delay times during the ablation process with a fluence 1.5 times the threshold fluence for each material. The color-coded regions represent distinct material phases—solid (dark blue), resolidified (bright blue), liquid (yellow), vapor (red). Only positive radius is shown, as the ideal surface has rotational symmetry. The time spans are selected to enable a phenomenological comparison of ablation behavior.

For both materials, CFD simulations with local fluence modulations due to surface roughness, as obtained from FDTD simulations with AFM measurements, were conducted. Results are given in Fig. 3, with equivalent color coding of the snapshots to Fig. 2. The exact fluence distribution from FDTD simulations is shown in Fig. 6 and will be discussed later.

For Al, the early frames at 2.075 ps and 11.575 ps reveal a prominent liquid phase adjacent to the irradiation zone, with locally varying vapor generation, markedly deviating from the ideal flat surface in Fig. 2. Even near the center, the material is not continuously vaporized. Additionally, the diameter of melted material is widened as compared to the ideal simulation yielding a width of 27 μm compared to 24 μm. The profile of the depth of vaporization shows a strong variation, not forming a continuous crater, yet the local depth of melting roughly follows the course of the initial surface topography. Notably, at 11.575 ps, vaporized material is ejected inhomogeneously from the surface, showing the spatially differing strongly thermodynamic conditions of the interaction zone. This trend continues as delay time extends to 101.575 ps and 201.575 ps. The vaporized material region is deepened, as observed with the ideal case, but local vapor pillars are propagating from the surface with differing velocities instead of a homogeneous vapor plume formation. At 1001.575 ps, the molten material is completely resolidified. Contrary, the polished AISI 304 surface shows the same ablation dynamic as observed in the ideal-surface case.

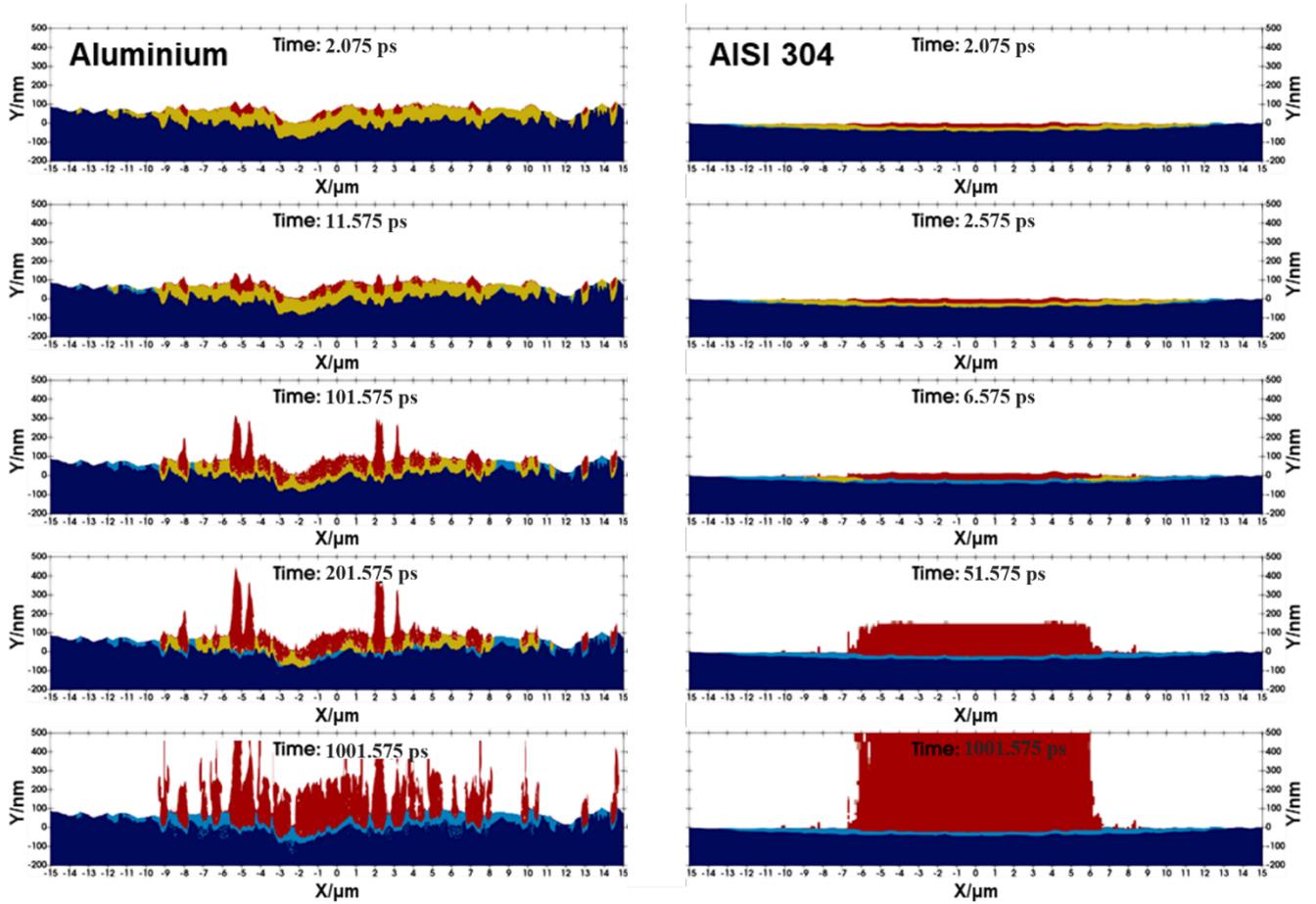

**Fig. 3** Sequential CFD simulation timeframes for Al and AISI 304 with 1.5 times the threshold fluence for each material, showcasing phase transitions on rough, measured surfaces over time. The color mapping indicates the material states—solid (dark blue), resolidified (bright blue), liquid (yellow), and vapor (red). The chosen delay times differ for both materials, as in Fig. 2.

Resolidification processes, as calculated with the simulation for both materials, are found starting already around 10 ps, which starkly contrasts the common picture of ultrafast melting, for which the resolidification starts in the ns regime [38,48]. Hence, resolidification is about 50 times faster in our case with a velocity of roughly 6 km/s as compared to expected order of magnitude of 100 m/s [11,49,50]. This may be related to the implementation of the phase transformations in the CFD model, as the TTM calculations indicate a liquid phase at this time. Nontheless, the observed resolidification velocities of ~6 km/s are in the order of magnitude of the sound velocity for both materials. Lastly, the velocity of the ejecting material of 3.3 km/s is in quite well agreement [51]. We attribute this behavior to the way the phase transitions have been implemented in the CFD solver. The phase is determined by the local pressure and the temperature, while the kinetics of the phase transition is mediated by the enthalpy of fusion and heat conduction only. This leaves out nucleation effects governing the phase changes by producing superheating and undercooling of the phases depending on the velocity of the interface of the phases, thus shifting the isotherm where phase transition occurs. This means, the vapor phase is already assigned to much bigger regions of the interaction zone as the velocity of the phase transition is not limited by superheating, as would be the case when the effect is included. Conceptually similar things can be stated about the solidification process and undercooling also leading to a significantly increased velocity of the resolidification. Thus, it becomes understandable why we observe a high amount of vapor at 2.075 ps and why the solidification is significantly faster than expected from literature. One encouraging sign for a compressible solver is, however, that the resolidification does not exceed the speed of sound, and the ejection speed of the vapor is in agreement with literature [51]. This kinetic of the phase changes can also influence the thermal conditions of the process zone, as different material properties are assigned to the cells depending on the phase composition present in the cell. The kinetics of the phase transformation and the associated assigned thermophysical properties warrant a close examination of the underlying mechanism and the way they could be implemented into the CFD solver.

The melted and vaporized region for Al is significantly deeper than compared to AISI 304. Regarding the time evolution, the ablation process of Al is characterized by a slower evolution compared to AISI 304. A notable plume generation on Al is observed at 101.575 ps, whereas AISI 304 exhibits similar phenomena already at 6.575 ps. Both lower depth and faster ablation dynamics for AISI 304 agree with the theory. AISI 304, a 3d-transition metal alloy, exhibits high collision frequencies due to localized 3d-band electrons near the Fermi edge, leading to reduced thermal conductivity and heightened electron-phonon coupling in comparison to Al, thus leading to a more confined energy distribution at the surface [52].

Concerning the real surface simulation, AISI 304 does not exhibit any differences compared to the idealized surface simulation, attributable to the low surface roughness of 2 nm. Contrary, Al exhibits significant local deviations, resulting in selective ablation processes on the rough surface as observed for e.g. modulated field distribution with surface plasmon polaritons [53]. This can be attributed to the near field scattering on the rough surface, which will be investigated in detail later. In conclusion, despite the lack of accurate resolidification kinetics, the influence of surface roughness on the crater morphology can be observed when compared to the ideal surface.

### *Predicted final state observables by TTM-CFD*

Crater diameter and ablation depths, as predicted by our simulations, are investigated in the case of an ideally flat surface. These final state observables were obtained by tracking the interface between resolidified material and vapor within the latest simulated time step in the CFD simulation (see, e.g., Fig. 2 blue region). Fig. 4(a) illustrates the resulting crater diameter in a *$D^2$ versus* $\log(\phi)$ representation, for both, Al (blue squares) and AISI 304 (red stars) at different incidence fluences. An initial inspection suggests for both materials a linear dependency, consistent with the heuristic *$D^2$* model [54] given by $D^2 = 2w_0^2 \ln(\phi/\phi_{\text{thr,D}})$. The fitting results (dashed lines) are summed up in Tab. 2, with a threshold fluence $\phi_{\text{thr,D}}$ for Al of 0.59 J/cm² and 0.13 J/cm² for AISI 304. For Al, the threshold fits the reported experimental result of 0.62(3) J/cm², determined by the same method [19]. For AISI 304, the simulation threshold deviates by a factor of two from the experimental value of 0.27(2) J/cm² [19].

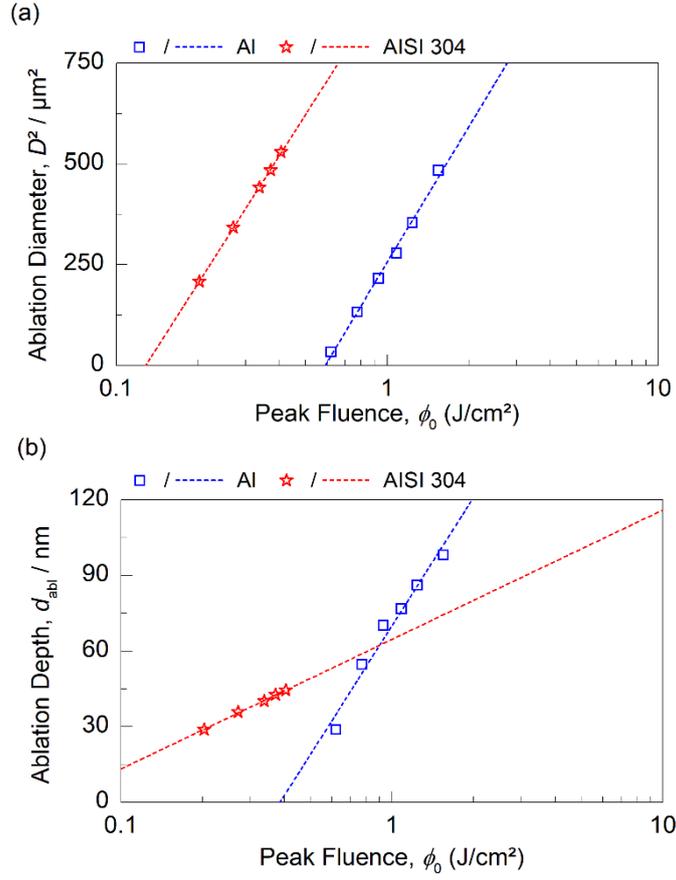

**Fig. 4** Simulation data and model fits of ablation diameter (a) and ablation depth (b) for Al (blue) and AISI 304 (red) depending on incident fluence. Fits of ablation diameter and ablation depth utilize the $D^2$-model and the $d_{\text{eff}}$-model, respectively.

Fig. 4(b) displays the simulated ablation depths from CFD at the center of the ablation crater ($x = 0$), plotted as depth $d$ versus $\log(\phi)$. The results correlate perfectly with linear absorption written as $d_{\text{abl}} = d_{\text{eff}} \ln(\phi/\phi_{\text{thr,d}})$, with the introduced $d_{\text{eff}}$ as the effective penetration depth [55]. Fitting, the resulting $d_{\text{eff}}$ and $\phi_{\text{thr,d}}$ yields 73.5 nm and 0.39 Jcm$^{-2}$ for Al, and for AISI 304 $d_{\text{eff}} = 22.3$ nm and $\phi_{\text{thr,d}} = 0.06$ J/cm$^2$. While AISI 304 $d_{\text{eff}}$ exhibits robust agreement with the experimental value of $d_{\text{eff}} = 15$ nm, Al considerably deviates by a factor of 2.6 from the experimental value of 28 nm. For $\phi_{\text{thr,d}}$, which are known to be lower than the $\phi_{\text{thr,D}}$ values [56], we find for AISI 304 a similar deviation as for the $D^2$-threshold with a factor of 3 compared to the experimental value of $\phi_{\text{thr,d}} = 0.18(2)$ J/cm$^2$ [19,52]. For Al, however, the experimental value yields 0.13 Jcm$^{-2}$, which contrasts the agreement between experiment and simulation for $\phi_{\text{thr,D}}$.

**Tab. 2** Final state results (upper row for each material) of diameter and depth and fitted values according to $D^2$-model and $d_{\text{eff}}$-model with Eqn. (1) and Eqn. (2), respectively. Lower row for each material shows the experimental results by Winter et al. [17].

|  |  | $\phi_{\text{thr,D}}$ (Jcm$^{-2}$) | $w_0$ (µm) | $\phi_{\text{thr,d}}$ (Jcm$^{-2}$) | $d_{\text{eff}}$ (nm) | $\eta_{\text{max}}$ (µm$^3$µJ$^{-1}$) | $\hat{\phi}$ |
|---|---|---|---|---|---|---|---|
| Simulation | Aluminum | 0.59 | 15.57 | 0.39 | 73.5 | 4.3 | 5.3 $\phi_{\text{thr}}$ |
| Experiment |  | 0.62(3) | 15.0(5) | ~0.13 | 28(2) | 3.0 | 3.5 $\phi_{\text{thr}}$ |
| Simulation | AISI 304 | 0.13 | 15.15 | 0.06 | 22.33 | 9.27 | 4.3 $\phi_{\text{thr}}$ |

| | | | | | | |
|---|---|---|---|---|---|---|
| Experiment | 0.27(2) | 15.0(5) | ~0.18 | 15(2) | 2.0 | 4.3 $\phi_{thr}$ |

Considering the ablation energetics, the ablation efficiency $\eta = V_{abl} / E_p$ is a crucial metric, where $V_{abl}$ represents the ablated volume and $E_p$ the incident pulse energy. $V_{abl}$ was evaluated by considering 2D-rotation symmetry using numerical radial integration of the extracted crater profile, according to the interface between resolidified material and vapor for the last simulated time step. We find for Al, at a fluence of 1.55 J/cm², a maximum with 4.3 µm³/µJ and for AISI 304 a value of 9.27 µm³/µJ at 0.37 J/cm². These normalized fluences $\hat{\phi}$ equal 5.3 $\phi_{thr}$ and 4.29 $\phi_{thr}$ for Al and AISI 304, respectively. Experiments by Winter *et al.* show ablation efficiency maxima of $\eta_{max} = 3.0$ µm³/µJ at a fluence of 3.5 $\phi_{thr}$ for Al and $\eta_{max} = 2.0$ µm³/µJ at a fluence of $\hat{\phi} = 4.3$ $\phi_{thr}$ for AISI 304. While for Al the efficiency value is in robust agreement with our simulation compared to experiments, for AISI 304 our simulations overestimate the experiments by a factor of 4.6. The fluence position of the maximum is however in perfect agreement for AISI 304, which is not the case for Al. The results and values from the experiment are summed up in Tab. 2. According to [48, 44], the ablation efficiency maximum may be estimated by the effective penetration depth and ablation threshold via $\hat{\eta} = 2d_{eff}/\hat{\phi}$. Thus, for AISI 304 with $\hat{\eta}_{sim}/\hat{\eta}_{exp} \approx 4.6$, the strong overestimation of ablation efficiency by the CFD results is mainly linked to the lower ablation threshold in the simulation by a factor of two. Calculating $(d_{eff,sim}\hat{\phi}_{exp})/(d_{eff,exp}\hat{\phi}_{sim})$ with the given quantities, we get a value of 2.9, which better aligns with the experiments. Therefore, it is evident that the ablation efficiency for AISI 304 in our simulations is higher than for Al, which is inconsistent with the experimentally determined data.

The results and deviations are briefly compared with simple TTM models for both materials. For Al, accurate predictions of fluence-dependent depth and diameter were obtained using a 1D-TTM with vaporization temperature as the ablation criterion and incorporating ballistic transport, which has not been considered in this work [57]. The simulation results for aluminum in this study successfully predict the diameter; however, they fail to accurately predict the depths. This discrepancy suggests that ballistic transport may play a non-negligible role. For AISI 304, similarly accurate results were achieved using a 2D-TTM simulation that accounted for the reflectance change as a function of electron temperature, a consideration absent in the present work, with the ablation criterion defined as 0.9 of the critical temperature [58]. The ablation criterion employed in this study is distinct from the criterion based on the evaporation temperature and might explain the difference.

Summarizing the energetic considerations and the final state observables, we can state that the predicted ablation metrics are in reasonable agreement with the experimental values and arising deviations are explained by appropriate analyses. Especially the perfect agreement between experiment and simulation of the *D²*-threshold for Al stands out. This is the case, although the CFD simulation does not incorporate the spallation mechanism explicitly while still yielding the same result as TTM-HD [23]. Therefore, despite the different kinetics of the phase transformations during ablation and solidification due to the implementation of the phase transformations in the CFD solver, the final state observables yield results comparable to those of experiments. Thus, the influence of surface roughness on simulation outcomes, as compared to idealized surfaces, can be discussed qualitatively. For this the derived $\phi_{thr,D}$ from the simulations are employed to adjust the incident fluences used in the simulations with implemented surface roughness. This adjustment is essential to achieve comparable thermal conditions between the simulations and the experiments, thereby enhancing the reliability of our comparison regarding the influence of surface roughness on the topography outcomes.

*Comparison with experiments*

We further compare the simulation results for polished real surfaces from AFM measurements with experiments taken from Winter et al. [19]. The comparison is made for 1.5- and 3.0-times ablation thresholds ablation as defined by the idealized simulation and experiments, respectively. This ensures comparability despite the differing absolute values regarding the resulting topography.

Fig. 5(a) and 5(b) illustrate the craters on Al and AISI 304, shown in blue and red, alongside the black lines representing the experimental data. For AISI 304, the experimental crater profile at 1.5 $\phi_{thr}$ presents a pronounced rectangular shape with steep sidewalls due to a minimal ablated depth created by the spallation mechanism, which is the expected process near the ablation threshold [19]. The simulation demonstrates commendable quantitative agreement with the experimental data, particularly for the smooth topography. However, the simulated crater profile is more parabolic, with a shorter plateau phase, overestimating the minimal ablation depth. With an increased fluence of 3.0 $\phi_{thr}$, approaching the efficiency maximum for single-pulse ablation of 4.3 $\phi_{thr}$ (simulation and experiment), both experimental and simulation results converge on a more parabolic crater shape, and again, the simulation perfectly predicts the surface topography of a rather smooth surface. The SEM images for 3.0 $\phi_{thr}$, as shown in Fig. 5(d), reveal a slight increase in surface roughness compared to the non-ablated surface area, separated from each other with a discernible crater edge.

In the case of Al, as depicted in the left panel in Fig. 5(a) in blue, the experimental results manifest (black solid line) a higher surface roughness when contrasted with AISI 304 for both experiment and simulation. At a fluence of 1.5 $\phi_{thr}$, the experimental ablation crater retains a discernible rectangular geometry, like that observed and explained for AISI 304 and pronounced sidewalls. The simulation tends to overestimate the surface roughness, suggesting spike-like features with an amplitude and lateral dimension of approximately 70 nm and 400 nm. Increasing the fluence to 3.0 $\phi_{thr}$, towards the efficiency maximum for single-pulse ablation at 5.3 $\phi_{thr}$ (simulation with ideal surface), the crater morphology for Al transitions to a more parabolic profile, analogous to the AISI 304 sample, with simulations and experimental outcomes displaying a greater congruence. The spike-like features maintain their visibility and dimensional characteristics, albeit with a reduced prominence. When contrasted with the SEM image of the crater on Al (3.0 $\phi_{thr}$), as illustrated in Fig 5(c), both the roughened surface texture and the definition of crater edges are evident, aligning with profile measurement data. With respect to these spike-like structures, the SEM image reveals the presence of approximately five nanostructures per 2 μm, hence, each with an average lateral dimension of about 400 nm. Consequently, despite overestimating the resulting surface roughness, simulations of real surfaces provide some predictive value concerning the surface topography.

## Discussion

The surface roughness influences the resulting crater morphology. To account for the initial surface topography prior laser irradiation, it is measured by AFM (Fig. 6), with blue color for Al and red for AISI 304. For reference, the ideal flat surface is depicted by black lines. The initial topography of Al with a $R_q$ surface roughness of 15 nm shows local deviations in depth from the ideal flat surface up to ±50 nm, contrary to the deviations of AISI 304 with ±5 nm ($R_q$ surface roughness of 2 nm). The resulting surface intensity distribution from FDTD simulations is shown in the second row of Fig. 6 with the same colors for the materials, as well as the ideal Gaussian intensity distribution with black dotted lines. For Al the field intensities deviate noticeably from the ideal Gaussian distribution (black dashed line) with a ratio of up to 3 between peak and minimum fluence whereby the peaks and minima are spaced periodically with ~400 nm over the entire crater surface. Even at a radius of 15 μm, a local intensity maximum is rivalling the original intensity peak of the ideal Gaussian distribution, resulting in an intensity increase ($I_{real}/I_{ideal}$) of ~7 times the intensity of the ideal Gaussian beam. The resulting crater of Al (Fig. 6) shows significantly increased surface roughness, when compared to the simulated ideal surface shown in black. The dimensions of ablated material regarding width and depth are comparable to the crater of the ideal surface as expected since we scaled the fluences to the derived threshold fluence from the ideal simulation. Notably, the local field intensity correlates with crater depth (Fig. 7), agreeing with the selective ablation processes observed in the transient analysis in the results section and thus creating a significantly rougher surface compared to the ideal surface. Hence, despite 15 nm $R_q$ surface roughness being two orders of magnitude below the laser wavelength, the impact of scattering significantly influences the resulting crater topography.

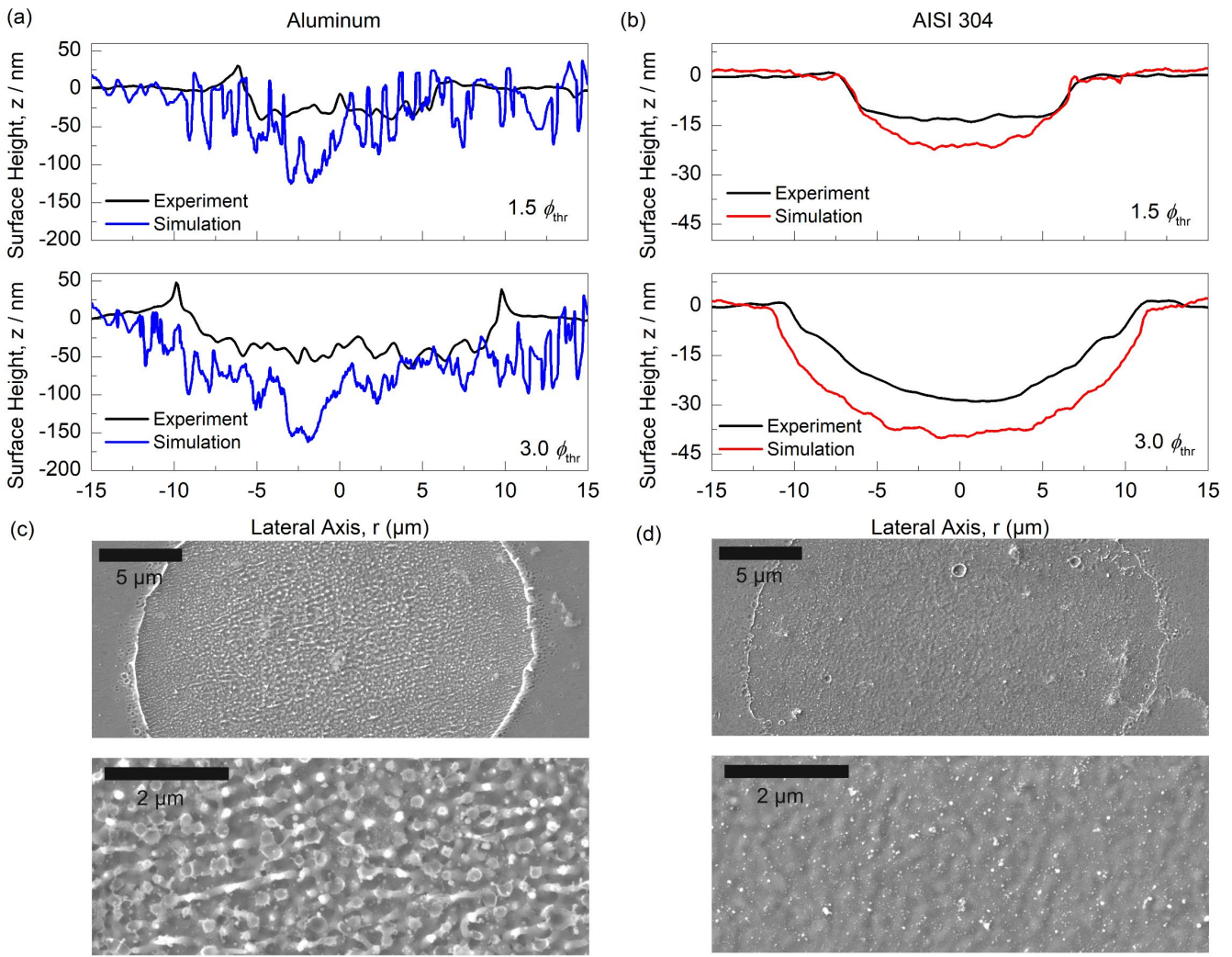

**Fig. 5** Panels (a) and (b) show crater cross-sections with the experimental profiles (black lines, from Winter *et al.* [17]) and simulation predictions (colored lines) at 1.5- and 3.0-times threshold fluence. Panels (c) and (d) provide SEM images at two magnifications for 3x threshold fluence, illustrating the real surface morphology.

On the other hand, the smoother surface roughness of AISI 304 with $R_q = 2$ nm result in negligible scattering effects, and therefore, the field intensities for AISI 304 adhere closely to the original Gaussian intensity distribution with deviations of less than ~2 % as depicted in Fig. 6(b). The resulting crater topographies for both cases, ideal and real, rough surface, are shown in Fig. 6(b) third row. The crater from the ideal simulation yields strong fluctuations at the crater rim, which could correspond to some local fulfillment and suppression of the evaporation condition within the ablation threshold. Yet, the quantization within the fluence catalogue of 0.5 mJ/cm² (absorbed energy), which is less than 1% of the ablation threshold, and therefore may not be attributed to this noisy behavior. For a comparison between the ideal and real case, the fluctuations at the crater rim of the ideal simulation are disregarded as they do not accurately reflect realistic results. The mean amplitude of fluctuations, estimated at approximately 8 nm, serves as an offset applied across the entire depth profile. To mitigate the impact of these fluctuations, any values surpassing the original surface level are set to zero, thereby flattening the surface profile fluctuations. With this adjustment, the modeled crater geometry and topography, shown in black in Fig. 6(b), are in precise alignment with the simulations of the actual surface.

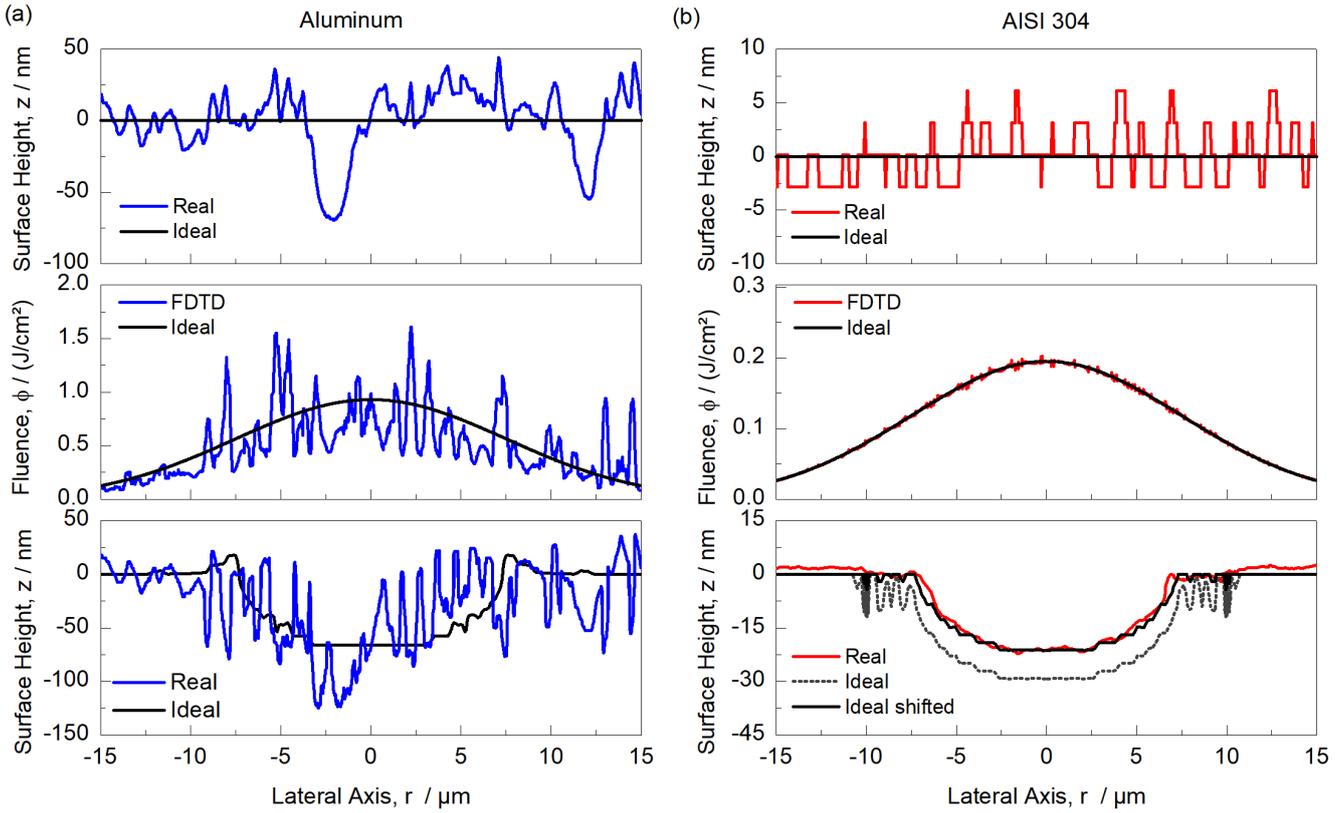

**Fig. 6** From top to bottom: the real rough surface profiles from AFM data, fluence distribution of FDTD simulation, and crater formation of CFD simulation with 1.5 times the threshold fluence as determined with the idealized simulations for (a) Al in blue and (b) AISI 304 in red, respectively. The black lines represent the idealized (smooth) surface conditions, serving as a reference for assessing the impact of surface roughness. Additionally, for AISI 304, the depth profile from the ideal simulation was adjusted to account for unphysical fluctuations.

The correlation between the local field intensity and the corresponding ablation depth will be discussed in the following to analyze lateral effects during and after the ablation process. A thorough experimental analysis in this regard would necessitate precise knowledge of the local laser field, which is currently beyond our reach, hence its exclusion.

First, the simulations with an ideally flat surface for Al, shown with black dots in Fig. 7(a, b) are analyzed. The course follows a steep increase at the threshold fluence of $\phi_{thr} = 0.59\,\text{J/cm}^2$ corresponding to a rectangular crater shape, as observed and explained in the results section. Local fluctuations are evident, which concur with expectations, considering the topography results of these simulations yielding surfaces with substantial roughness, as previously noted. In the ideal case, neither an inhomogeneous fluence distribution nor strong lateral heat transport is expected due to the ideal Gaussian distribution. In combination with the observation that material is being transported to the crater edge, as seen in the simulated transient ablation dynamics in form of swollen crater edges, we conclude that melt flow is the driving factor for roughness generation despite idealized conditions. This can be understood with temperature modulation, temperature-dependent surface tension and the interaction of the involved fluids. Additionally, to the lateral fluid motion due to the pressure gradient of a fluence-dependent ablation, the temperature towards the crater edge decreases, the surface tension of the molten material in turn increases, resulting in an additional force inducing lateral melt flow from the center towards the crater's edge [59,60].

For the real surface simulations of Al, the local ablation depth as compared to the incident Gaussian fluence distribution depicted in Fig. 7(a), shows no correlation up to 1 Jcm$^{-2}$ and a rough correlation for higher fluences. The comparison of the local ablation depth to the local fluence distribution as determined by FDTD (Fig. 7(b)), however, shows a clear correlation with the ablated depths, yet significantly stronger fluctuations as compared to the ideal case are evident. The fluctuations in the real case in Fig. 7(a) encompass optical scattering effects that modulate local fluence, along with thermophysical influences and melt flow. Conversely, the latter case in Fig. 7(b) is characterized solely by thermophysical influences and melt flow induced by pressure gradients.

The physical consequences induced by inhomogeneous energy deposition are far-reaching and can explain the observed extent of the fluctuations. The high local fluence variations may lead to strong lateral heat conduction during the TTM simulation with 1D approximation. As electrons are the primary factor for heat conduction during pulse irradiation, this effect can be estimated by assuming a linear relationship between electron temperature $T_e$ and electric field intensity $T_e(x) \propto I(x)$. The resulting heat conduction is then directly connected to the intensity distribution as $\dot{Q}_e \propto \Delta I / \Delta x$. For instance, the field distribution in Fig. 6 (a and b, second row) shows local maxima at approximately $r \approx +2.5$ µm, about 3 times higher than the nearby minima located roughly 400 nm away. By contrast, the maximum variation in an ideal Gaussian intensity distribution over a 400 nm span in the vicinity of the highest slope at r ≈ 7.5 µm is around 10%. Therefore, in this case, the lateral heat transport at $r \approx +2.5$ µm could be approximately 30 times higher than that on an ideal surface, reducing the ratio to vertical heat transport to only a factor of 30, contrary to the initially estimated factor of $10^3$. Furthermore, the approximation of constant refractive index might induce notable changes as well. With increased fluence, the reflectivity drops as known from experiments [23], thereby increasing the local absorptance, consequently amplifying the lateral electron temperature gradient. Although the TTM simulation accounts for temperature dependent variations of the complex refractive index by incorporating changes in the dielectric function and density, the FDTD does not. Hence, during pulse irradiation local absorptance changes might notably influence the resulting absorbed field distribution. For a quantitative assessment, a detailed 2D-TTM simulation, including surface scattering effects, is required. Nevertheless, the local fluence variations induced by surface roughness challenge the assumption that lateral transport effects can be neglected during the first 1.575 ps by critically discussing the overestimation of surface roughness in this case and highlighting the suppressed heat transport hindering a smoothening of the optically induces temperature modulation before ablation takes place. The underestimation of lateral heat transfer would result in increased spatial temperature, pressure and surface tension variation. The temperature variation increases the driving force for Marangoni convection. Meanwhile, temperature and pressure modulations are creating a more variable ablation dynamic due to higher local pressure gradients (evident by locally varying ejection velocity of vapor), leading to melt flow away from high-pressure regions. Furthermore, the locally varying pressure changes the melting/ boiling temperatures and, therefore, leads to a different ablation depth even for equal local fluences, increasing the fluctuations without lateral material flow effects even being necessary. As seen in the transient ablation dynamics the resolidification velocity is 50 times faster than expected from simulations [11]. This constrains the liquid behavior in the ablation zone by prematurely freezing the otherwise developing fluid motion and prohibiting the examination of the final state of crater morphology. However, we can assume that the acting surface tension would further smoothen the surface roughness observed in the case of the simulated real Al surface in Fig. 5(a).

For AISI 304, the ideal-surface simulation was adjusted regarding the unphysical fluctuations at the crater rim, as explained before. The course follows a steep increase in ablation depth at the threshold fluence of 0.13 mJ/cm², which corresponds to a minimal ablation depth and a rectangular crater shape as discussed for Al in the previous section. Concerning the comparison between ideal and real surface simulations, the correlation between ablation depth and local fluence perfectly align with each other when acknowledging the unphysical behavior at the crater's edge. Thus, non-local effects, as observed in the case of Al, are not pronounced since fluence variations are negligible for AISI 304. The observed depth fluctuations are not due to optical scattering as they are also apparent for the ideal case. The fluctuations for the ideal case of AISI 304 are not as pronounced as for Al. Without fluence modulation and therefore negligible lateral heat transport, this may be attributed solely to material-specific properties, as surface tension and density of the steel are significantly higher, therefore showing a higher driving force for smoothing out local variations in melt surface height while resisting lateral pressure driven melt flows via inertia.

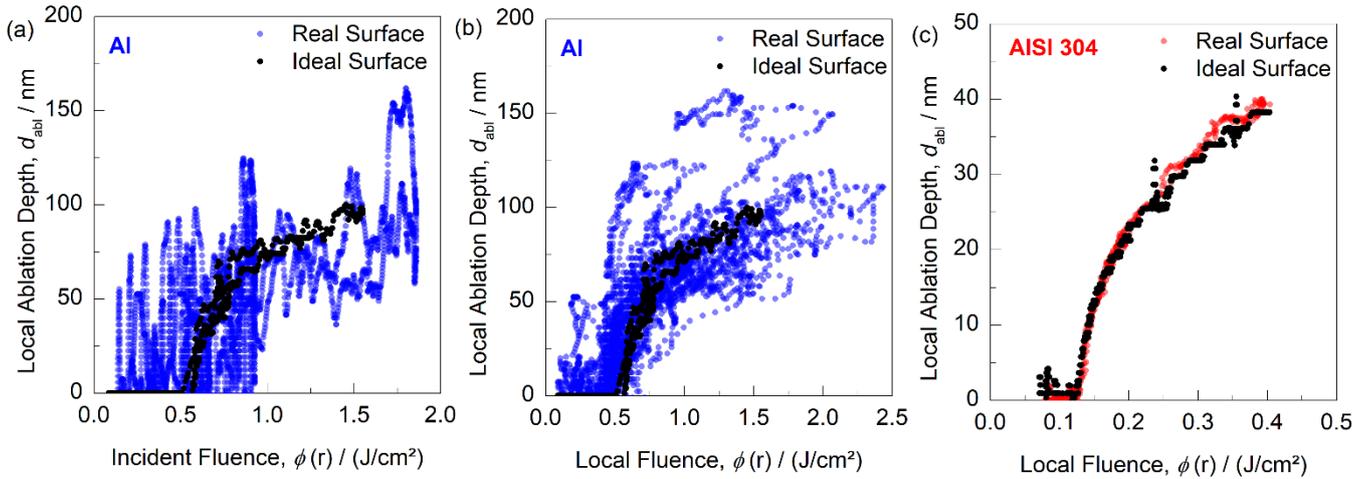

**Fig. 7** Correlation between the ablation depth and local fluence: (a) Gaussian distribution vs. local ablation depths for Al and (b) with the used FDTD derived local fluences vs. local ablation depth. (c) shows the FDTD derived fluence distribution vs. the local ablation depth. Colored dots represent the simulation results of real surface; black dots show ideal surface predictions.

## Summary and Conclusion

In this study, a multi-scale simulation framework for USP laser ablation was developed to analyze the impact of the surface roughness of real specimens of Aluminum and steel (AISI 304) on the resulting crater morphology and ablation dynamics for USP laser ablation. The materials were selected due to their strongly differing material properties. The framework provides the possibility of a detailed examination across the relevant timescales of the ablation process, from the sub-picosecond to the nanosecond range, by integrating FDTD simulations for receiving the local fluence due to surface roughness, EOS to determine the thermodynamic properties of the involved phases, the TTM for exciting the material to its initial state and a compressible multi-phase CFD model for relaxing the initial state.

The final state observables of the simulations (fluence-dependent depths and diameters) aligned well with experimental observations. Despite quantitative deviations, the simulation model, for example, successfully predicted a $D^2$-threshold fluence of 0.59 Jcm$^{-2}$ for Al (experiment 0.62 Jcm$^{-2}$) and a reasonable value for AISI 304 of 0.13 Jcm$^{-2}$ (experiment 0.27 Jcm$^{-2}$). The observed velocities of the phase transitions, however, differ from the expected behavior from literature. These observations highlight the necessity of explicitly including effects associated with superheating and undercooling resulting from the high velocities of the phase transformations in laser ablation to fully describe ablation dynamics and predict the surface topography in detail. However, further increasing model complexity currently does not seem feasible due to already prohibitive computational cost, limited scaling behavior and challenging numerical stability in the compressible multiphase CFD model.

In the case of Al with a surface roughness of 15 nm (laser wavelength 1056 nm), simulations showed pronounced optical scattering effects, leading to selective local ablation processes resulting in a significantly altered crater topography compared to the ideal case. This finding challenges the 1D-TTM approximation as local intensity modulations induce possible lateral effects in the order of magnitude of axial heat transfer. Furthermore, it was shown that the outcomes of large-scale simulations and their concordance with experimental results are significantly influenced by surface roughness, even with $R_q$ values ~2 orders of magnitude below the laser wavelength, underscoring that simulations predicated on idealized surface conditions cannot yield perfect results for some materials. Hence, we highlight the importance of incorporating material-specific initial surface topography for laser ablation simulations.

Future research should investigate varying surface conditions of the material to quantify the impact of surface roughness and to determine the threshold at which incorporating surface roughness into ablation simulations becomes essential. The observed effects on the surface are known to be important for multi-pulse ablation, as they repeatedly alter the surface and thus introduce increasing scattering effects. Understanding these interactions is essential for the successful simulation of LIPSS formation and for predictably creating or preventing LIPSS on metals.

**Author contributions**

Conceptualization: DR, MD, HPH and MSc; methodology: NT, MSt, MD and DR; formal analysis and investigation: NT, MSt, MD, DR, HPH and MSc; FDTD simulation: NT and DR; TTM-CM calculations: DR and JV; TTM-CFD simulation: MSt; writing - original draft preparation: NT, MD, MSt and DR; writing - review and editing: NT, MD, MSt, DR, HPH and MSc; funding acquisition: HPH and MSc; resources: HPH and MSc; supervision: HPH and MSc.

**Funding**

This work was supported by Deutsche Forschungsgemeinschaft under Grant 428973857 and the project MEBIOSYS, funded as project No. CZ.02.01.01/00/22_008/0004634 by Programme Johannes Amos Commenius, call Excellent Research.

The authors gratefully acknowledge funding of the Erlangen Graduate School in Advanced Optical Technologies (SAOT) by the Bavarian State Ministry for Science and Art.

**Competing Interests**

The authors declare that there are no conflicts of interest regarding the publication of this paper.

**Data availability**

The datasets generated and analyzed during the current study are available from the corresponding author on reasonable request.

**Appendix**

    A. Additional information about the FDTD simulations

The software used for FDTD simulations is Ansys, Lumerical 2023 R1.3. Fig. 8 shows the software interface, depicting the simulation setup for Al in this case. The simulation region, outlined in orange, comprises Yee cells with a defined resolution in a 2D space. The boundaries are set to perfectly absorbing layers with 8 layers, effectively mitigating any reflectance. The simulation time is configured to be six times the pulse width of the source, and the temperature is maintained at 300 K. The width of the simulation region is set to six times the pulse width of the Gaussian source. The Gaussian source is depicted at the bottom. The purple arrow indicates the propagation direction, while the blue arrow denotes the polarization of the electric field.

The yellow regions represent field monitors that calculate the electric fields in the frequency domain for each cell. The material shown in red is characterized by a complex refractive index, with its surface topography implemented using AFM data in the form of a text file. Field monitors in the vicinity of the vacuum-material interface enable index determination, facilitating the definition of a mask. This configuration provides a straightforward method to calculate the fields at the cells located on the surface adjoining the vacuum. Additionally, monitors were positioned in front and behind the Gaussian source to measure the total reflected light and verify the energy conversation with absorption calculation done with the electrical fields in the material. The latter can be done with a finished script provided by Ansys. The results yielded a total reflectance of 92.4% and absorptance of 7% resulting in a total energy of 98.4%. The remaining 1.6% are likely lost at the boundaries, as the rough surface increases diffuse scattering and hence, portions of the light reaching the boundaries prior to detection by the monitors may be absorbed.

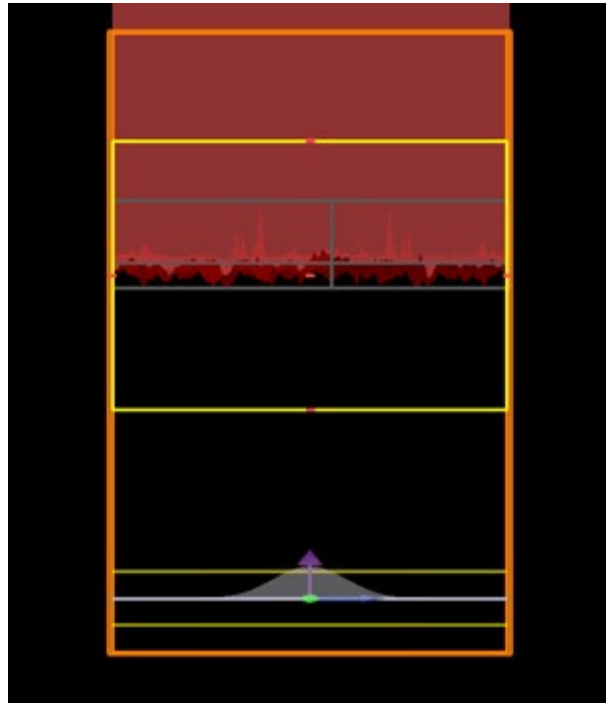

**Fig. 8:** Simulation setup in Ansys Lumerical 2023 R1.3 for FDTD of Al, showing the simulation region (orange), field monitors (yellow), and material with complex refractive index (red). The Gaussian source and its propagation and polarization directions are indicated by the purple and blue arrows, respectively.

Due to limited computational resources, the simulation was performed in two dimensions. To assess the differences compared to a full 3D simulation, a comparative analysis was conducted. For this comparison, the simulation region was reduced to 20 µm, with a Gaussian pulse of 5 µm diameter and amplitude of 1 V/m. The resolution was set to 50 nm laterally and 5 nm along the propagation direction. The results for the surface fields determined with the method explained before for the cases of 2D and 3D simulations are presented in Fig. 9. Minor differences are observed possibly due to diffraction effects introduced by the additional dimension. Nonetheless, the field distribution is well preserved in the 2D approximation.

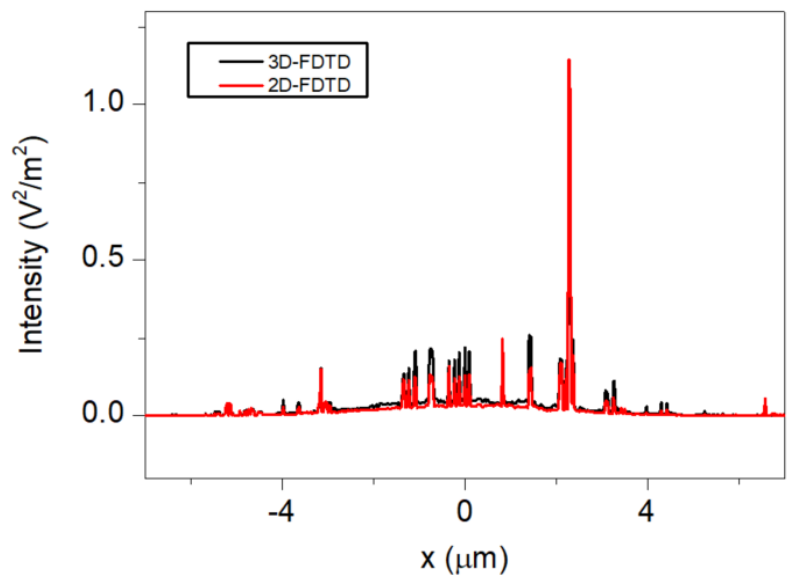

**Fig 9**: Intensity distribution comparison between 2D (red) and 3D (black) FDTD simulations. The plot shows the field intensity on the surface as a function of lateral position. While slight variations arise from diffraction in the third dimension, the overall field distribution remains consistent between the two simulations.

B. Fitting for dynamic viscosity

The polynomial fitting of the dynamic viscosity is given by

$$\mu_{Al} = 9.611416847138^{-3} - 1.968496706832412^{-5} \cdot T + 1.745827792206794^{-8} \cdot T^2 - 7.87503326214483^{-12} \cdot T^3 + 1.778951802353309^{-15} \cdot T^4$$

for aluminum and by

$$\mu_{AISI\ 304} = 6.1^{-2} - 5.7748^{-5} \cdot T + 1.9319^{-8} \cdot T^2 - 2.2054^{-12} \cdot T^3$$

for AISI 304 throughout the liquid state (in Pa*s).

C. Equation of States for AISI 304

The equation of states for AISI 304 (Fe70Cr19Ni11) were derived based on [28]. The results are shown in Fig. 10.

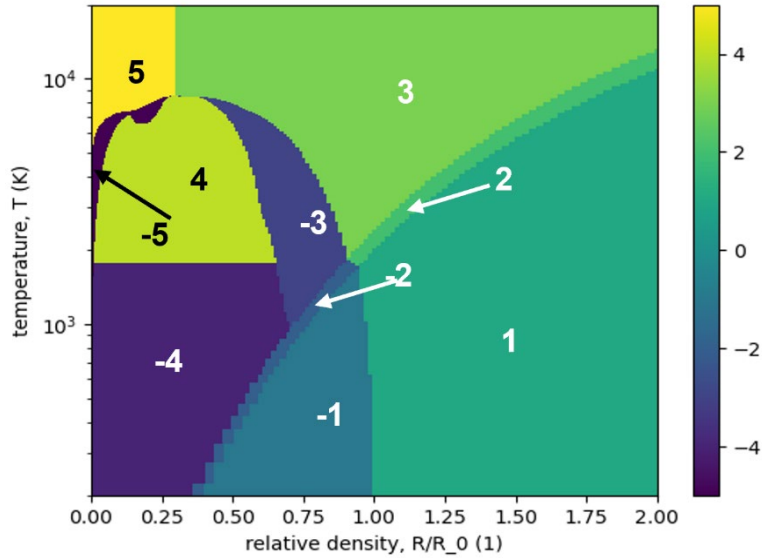

**Fig. 10:** The phases are defined as follows: 0 represents void, 1 denotes solid, 2 corresponds to a solid-liquid mixture, 3 refers to the liquid phase, 4 indicates a liquid-gas mixture, and 5 describes the gas phase. A minus sign preceding the phase number indicates its metastable counterpart.